\documentclass{svproc}

\usepackage{amssymb,amsmath}
\usepackage{amsfonts}

\usepackage[misc,geometry]{ifsym}  
\usepackage{hyphenat}              
\usepackage{subeqnarray}
\usepackage{enumitem} 

\usepackage{xcolor}
\usepackage{graphicx}

\usepackage{url}              

\newcommand{\coo}{CO\textsubscript{2}}

\newcommand\inlineeqno{\stepcounter{equation}\ (\theequation)}

\newcommand{\dpar}[2]{\frac{\partial #1}{\partial #2}}
\newcommand{\tmdder}[3]{\left(\frac{\partial #1}{\partial #2}\right)_{\!\!#3}}

\newcommand{\Pt}{\widetilde{P}}
\newcommand{\Dim}[1]{\widetilde{#1}}
\newcommand{\Ref}{_{\mathrm{r}}}

\newcommand{\tnp}{^{n+1}}
\newcommand{\tns}{^*}
\newcommand{\tn}{^n}
\newcommand{\jmh}{_{j-\frac{1}{2}}}
\newcommand{\jph}{_{j+\frac{1}{2}}}

\begin{document}
\mainmatter              
\title{Non-equilibrium Model for Weakly Compressible Multi-component Flows:\\ the Hyperbolic Operator}
\titlerunning{Weakly Compressible Multi-component Flows: the Hyperbolic Operator} 
\author{Barbara Re \textsuperscript{\Letter} \and R\'{e}mi Abgrall}
\authorrunning{B. Re, R. Abgrall} 

\tocauthor{Barbara Re, and R\'{e}mi Abgrall}
\institute{Universit\"{a} Z\"{u}rich, Institute of Mathematics, Z\"{urich}, CH8057, Switzerland,\\
\email{barbara.re@math.uzh.ch}}

\maketitle              

\begin{abstract}
We present a novel pressure-based method for weakly compressible multiphase flows, based on a non-equilibrium 
Baer and Nunziato\hyp{}type model.
In this work, we describe the hyperbolic operator, thus we do not consider relaxation terms.
The acoustic part of the governing equations is treated implicitly to avoid the severe restriction on the time step imposed by the CFL condition at low-Mach.
Particular care is taken to discretize the non-conservative terms to avoid spurious oscillations across multi-material interfaces.
The absence of oscillations and the agreement with analytical or published solutions is demonstrated in simplified test cases, which confirm the validity of the proposed approach as a building block on which developing more accurate and comprehensive methods.
\end{abstract}
\section{Introduction}
Weakly compressible multiphase (or multi-component) flows may arise in several applications,
as for instance the transport of \coo\ for carbon capture and storage (CCS)~\cite{Munkejord2015},
in incidental configurations of water nuclear power plants~\cite{Coquel2017}, or in additive manufacturing.
The former application, in particular, has motivated the present work.
In normal-operating conditions, the transport pipelines are primary designed to contain liquid or super-critical \coo,
but multiphase flows may occur due to fluctuating \coo\ supply or during transient events~\cite{Munkejord2015}.
The transported flow may contain also diverse impurities, whose state is not necessarily the same of the main component.
In this regime, although the velocity is considerably smaller than the speed of sound, i.e. the Mach number is very low, compressibility effects cannot be neglected.
The design, safety assessment and performance analysis of these and similar applications could undoubtedly benefit from efficient and robust computation fluid dynamics (CFD) tools.

From the numerical point of view, the standard schemes widely used for inviscid compressible flows fields containing shock waves, contact discontinuities, and strong rarefactions, are inadequate for weakly compressible flows, even single-phase ones.
The causes of this failure, usually referred to as low-Mach limit, are multiple.
First, the use of an explicit time integration scheme increases dramatically the computational time.
Indeed, at low Mach, the problem is stiff and the acoustic effects result in a severe time step restriction imposed by the CFL condition~\cite{Wenneker2002}.
A further difficulty concerns accuracy. 
Generally, approaching the limit $M = 0$, the numerical solution of the compressible Euler equations does not converge to solutions satisfying the equations for incompressible flows, except for a restricted class of special initial data (the so-called well-prepared case).
In particular, when a flux difference splitting method is used,
the discrete scheme cannot recover the correct scaling of the pressure with the Mach number and this error increases as the Mach number goes to zero~\cite{Guillard1999}.

The low-Mach limit has been investigated since long time in single-phase CFD and the literature presents several contributions on how to develop a numerical method that provides accuracy and efficiency for weakly compressible flows.
The proposed answers can be grouped roughly into two main strategies.
The first one improves the low-Mach behavior of compressible schemes by means of preconditioning or artificial viscosity~\cite{Guillard2004}. This approach leads to problem-specific solutions, which can be hardly generalized, and increases the computational burden.
Alternatively, compressibility can be introduced into the incompressible schemes, by treating implicitly the acoustics and circumventing the issues related to the stiffness of the governing equations~\cite{Xiao2004}.
This latter research field was pioneered by Harlow and Amsden~\cite{Harlow1968,Harlow1971} and it has represented the core of several recent and outgoing research activities aimed at developing Mach-uniform schemes, e.g., \cite{Bijl1998,Munz2003,Kwatra2009,Ventosa-Molina2017}.
In this work, we exploit some of the techniques proposed within this branch of research to build a robust and flexible method for weakly compressible, non-equilibrium multiphase flows.

While conceiving a CFD tool for multiphase flows, two of the main challenges that we have to handle are the multiscale nature of the field and the presence of the dynamic interfaces that separate distinct fluids (or phases).
These issues can be addressed in numerous ways: among them, we can distinguish interface-tracking methods   and interface capturing methods.
The former ones, as for instance level-set, volume-of-fluid, front-tracking, and ad hoc arbitrary Lagrangian-Eulerian methods, fully resolve the interfaces and smooth out the fluid properties across them. 
The latter ones dynamically capture the interface as part of the numerical solution, by treating each fluid as a separate continuous. These methods are also known as multi-fluid or diffuse interface method~\cite{Saurel2018}.
This work focuses on this second class, which is more suitable to deal with the dynamic creation of interfaces. More specifically, we adopt a Baer and Nunziato (BN)-type model~\cite{Baer1986} (also known as 7-equation), which copes with the multitude of scales characterizing multiphase flows by means of averaging procedures.
Assuming non-equilibrium of phasic pressures, velocities and internal energies, it allows the most general description of multiphase flows and the description of each component through its own thermodynamic model. 

Some research activities has been already devoted to weakly compressible multiphase flows, but most of them focus on simplified BN-type models in which equilibrium in the pressure~\cite{Pandare2018} and also in the velocity and the temperature~\cite{Bernard2014} is imposed.
On the contrary, Coquel et al.~\cite{Coquel2017} adopt a full non-equilibrium BN-type model, but they make the simplifying assumption of isentropic flows.
A part from the broader generality, a nice property that motivates the use of BN model is the hyberbolicity, which is not guaranteed by the 6-equation one-pressure model.
In this work, we work with the symmetric variant proposed by Saurel and Abgrall~\cite{Saurel1999}, which includes pressure and velocity relaxation terms to model how the phasic equilibrium is reached at the interface.

The goal of the paper is to present a preliminary 1D version of a novel numerical method for weakly compressible, non-equilibrium multiphase flows.
Accordingly, we present how we have derived the model and the numerical discretization in Sec.~\ref{s:method}.
Particular care has been devoted to the discretization of the non-conservative terms involving the gradient of the volume fraction, which couple the evolution of the different phases.
Then, in Sec.~\ref{s:results}, we verify its validity in simplified test cases, aimed at recovering the single-phase behavior and verifying the correct discretization of non-conservative terms, i.e. the absence of spurious oscillations across multi-material interfaces.
Finally, Sec.~\ref{s:concl} draws the conclusions and outlines the future steps towards the long-term work.

\section{Numerical Method}
\label{s:method}
A winning strategy to attain an efficient numerical method for weakly compressible flows consists in approximating the sonic terms, which in the low-Mach limit are associated with an almost infinite sound propagation rate, in an implicit fashion.
Typically, the methods that pursue this strategy for single-phase flows exploit a staggered description of the flow variables to avoid spurious pressure oscillations, and they solve the governing equations in a segregate way by updating the variables to the next time level through intermediate sub-steps.
In low-Mach regimes, pressure-based methods are considered more suitable than density-based ones, because of the weak coupling between these two variables. Indeed, the density is approximately constant, while the pressure not, so if density is used as primary variable, small errors on it results in large errors in the pressure.

In the followings, we illustrate how we have conceived a pressure-based, segregated strategy for non-equilibrium multiphase flows.

\subsection{The continuous model}
\label{ss:model}
The first key ingredient of the proposed model is a pressure decomposition, which is required to retrieve the correct order of pressure fluctuations and to converge to correct approximation of the incompressible Euler equations~\cite{Wenneker2002}.
We replicate here the choice made in~\cite{Bijl1998} and we define the dimensionless pressure as
\begin{equation}
P = \frac{\Pt - \Pt\Ref}{\rho\Ref u\Ref^2} \,.
\label{eq:Pscaling}
\end{equation}
where $\rho$ is the density, $u$ the velocity, and the subscript $\mathrm{r}$ and the accent $\widetilde{\cdot}$ indicate reference and dimensional quantities, respectively.
Implicitly, this scaling filters out the acoustics, as it postulates an asymptotic expansion of the pressure as
$$P = P^{(0)} + M\Ref^2 P^{(1)} + \mathcal{O}(M\Ref^3)
\qquad \mathrm{with} \quad 
M\Ref^2 = \frac{\Dim{\rho}\Ref \Dim{u}\Ref^2}{\Pt\Ref}\,$$
where $M\Ref$ is a reference Mach number expressing the overall compressibility of the flow field.
We can interpret this scalding also as a decomposition between the thermodynamic pressure ($\Pt\Ref$) and a component that satisfies the momentum equation~\cite{Wenneker2002,Munz2003}.

We want to apply this scaling to the Baer and Nunziato model. As told before, we work with the symmetric variant proposed by Saurel and Abgrall~\cite{Saurel1999},
which gives the possibility to use the same equations and the same numerical methods at all computational cells.
In this preliminary work, we focus on the hyperbolic part, so we do not consider the relaxation terms.
Consequently, considering only 2 phases (denoted with subscripts $i$ and $i^*$), the dimensional form of the governing equations in 1D is
\begin{subeqnarray}
\label{e:dim}
\dpar{\alpha_i}{\Dim{t}} + \Dim{u}_I \dpar{\alpha_i}{\Dim{x}} &=& 0\slabel{se:dim_a}\\[1ex]
 \dpar{(\alpha_i \Dim{\rho}_i)}{\Dim{t}}  + \dpar{(\alpha_i \Dim{m}_i )}{\Dim{x}} &=& 0  \slabel{se:dim_d}\\
 \dpar{(\alpha_i \Dim{m}_i )}{\Dim{t}}  + \dpar{(\alpha_i \Dim{m}_i \Dim{u}_i  + \alpha_i \Pt_i)}{\Dim{x}} &=&
   \Pt_I \dpar{\alpha_i}{\Dim{x}}\slabel{se:dim_m}\\
\dpar{(\alpha_i \Dim{E}_i)}{\Dim{t}}  + \dpar{\left(\alpha_i ( \Dim{E}_i + \Pt_i) \Dim{m}_i/\Dim{\rho}_i \right)}{\Dim{x}} &=&
   \Pt_I \Dim{u}_I \dpar{\alpha_i}{\Dim{x}} \slabel{se:dim_e}
\end{subeqnarray}
where the last three equations are repeated also for the phase $i^*$.
This system is written for the volume fraction $\alpha$ and for the conservative variables $\rho$, $m$, and $E=e + m^2 /(2 \rho)$ of each phase.
In addition to the standard Euler equations, the system includes transport terms that stem from the averaging process that underlies BN-type model. To close the system, we have to define the interfacial pressure and velocity ($P_I$ and $u_I$).
Following~\cite{Saurel1999}, we define them as the weighted averages among the phases,
i.e. $P_I = \sum_i \alpha_i P_i$ and $u_I = \sum_i \alpha_i m_i / (\alpha_i \rho_i)$, although different alternatives are available in the literature.
The system~\eqref{e:dim} has to be complemented by a thermodynamic model for each fluid.
In this preliminary work, we use the stiffened gas model, but we leave the formulation as general as possible, to be able to easily add more accurate equations of state in the future.

Since our goal is the development of a pressure-based method, we need to formulate the governing equations in primitive variables instead of conservative ones. 
We are aware of the possible problems that may arise from working with a non-conservative formulation, but at the moment our targets are low-Mach flows, so not experiencing strong shock waves; then a local correction is sufficient to converge to the correct weak solution. Recently, this idea has been successfully applied to multiphase flows modeled according to the simplified 5-equations model~\cite{Abgrall2017}.
Therefore, we substitute Eq~\eqref{se:dim_e} with the following:
\begin{equation}
\label{e:dim_P}
\alpha_i\dpar{\Pt_i}{\Dim{t}} + \alpha \Dim{u}_i \dpar{\Pt_i}{\Dim{x}} 
 + \alpha_i \Dim{\rho}_i \Dim{c}_i^2  \dpar{\Dim{u}_i}{\Dim{x}}  =
\Dim{\rho}_i \Dim{c}_{i,I}^2  (\Dim{u}_I - \Dim{u}_i)  \dpar{\alpha_i}{\Dim{x}} 
\end{equation}
where ${c}_i$ is the standard speed of sound of phase $i$ and $ c_{i,I}$ is a kind of interfacial speed of sound, which we have defined  as
$ c_{i,I}^2 = \chi_i + \kappa_i \frac{P_I + e_i}{\rho_i} \,,$ with
$ \chi = \tmdder{P}{\rho}{e}$ and $\kappa = \tmdder{P}{e}{\rho}$.
We highlight that this model, as all 7-equation BN-type ones, does not require to define a speed of sound for the mixture, which can be considered as an open-issue in multiphase research.
 
The final step to formulate the target model is to make dimensionless the BN-type model given by \eqref{se:dim_a}--\eqref{se:dim_m}, and \eqref{e:dim_P} according to the scaling given by~\eqref{eq:Pscaling}.
The variables that do not involve pressure are scaled as usual, according to the reference density, velocity, and length ($L\Ref$).
On the other hand, the scaling of the thermodynamic variables, and in particular of the speed of sound, requires more care. Inserting Eq.~\eqref{eq:Pscaling} into the definition of $\Dim{c}^2_i$, we have 
\begin{equation}
\Dim{c}^2 =
\left[\chi + \kappa \frac{P + e}{\rho}\right] \Dim{u}\Ref^2 + \kappa \frac{\Dim{P\Ref}}{\rho \Dim{\rho}\Ref}=
\left[c^2 + \kappa \frac{P\Ref}{\rho}\right] \Dim{u}\Ref^2
\end{equation}
where $c^2 = \chi + \kappa \frac{P + e}{\rho}$ has the same expression of the dimensional speed of sound, and ${P\Ref} =\Dim{P}\Ref / (\Dim{u}\Ref^2 \Dim{\rho}\Ref$  is the dimensionless reference pressure.
A similar result is obtained also for $\Dim{c}^2_{i,I}$.

Omitting all the passages for brevity, the dimensionless formulation reads
\begin{subeqnarray}
\label{e:adim}
\dpar{\alpha_i}{t} + u_I \dpar{\alpha_i}{x} &=& 0\slabel{se:adim_a}\\[1ex]
 \dpar{(\alpha_i \rho_i)}{t}  + \dpar{(\alpha_i m_i )}{x} &=& 0  \slabel{se:adim_d}\\
 \dpar{(\alpha_i m_i )}{t}  + \dpar{(\alpha_i m_i u_i  + \alpha_i P_i)}{x} &=&
   P_I \dpar{\alpha_i}{x}\slabel{se:adim_m}\\
M\Ref^2 \left[\alpha_i \dpar{P_i}{t} + \alpha_i u_i \dpar{P_i}{x} 
 + \alpha_i \rho_i c_i^2   \dpar{u_i}{x} \right] &=&
M\Ref^2  \rho_i c_{i,I}^2  \left[ (u_I - u_i) \dpar{\alpha_i}{x} \right] \nonumber\\
  &+& \kappa_i \bigg[u_I \dpar{\alpha_i}{x}  - \dpar{(\alpha_i u_i)}{x} \bigg]\slabel{se:adim_P} \;.
\end{subeqnarray}
We can observe that the first three equations are identical to their dimensional counterparts, while the pressure equation has a special expression, thanks to which, as $M\Ref \rightarrow 0$, we have
$\dpar{\alpha_i}{t} +\dpar{(\alpha_i u_i)}{x} = 0$.
This can be viewed as the multiphase counterpart of the kinematic constraint for the incompressible single phase flows $\vec{\nabla} \cdot \vec{u} = 0$.

\subsection{The discretization}
\label{ss:discretization}
As common in low-Mach schemes for single phase flows,
we exploit a staggered description of the flow variables to avoid spurious pressure oscillations. This means that the scalar, thermodynamic variables are stored are the cell centered, while the vectorial, velocity-related quantities are stored at the cell faces.
Under this configuration, it is natural to solve the governing equations in a segregate way, as the computational cells for the momentum equation are different from the ones for the density and energy equation.

\subsubsection{The semi-implicit temporal discretization.}
The solution at time level $t\tnp$ is computed through intermediate sub-steps in which the convective velocity is approximated explicitly.
Moreover, the computation of the momentum is split into two sub-steps: first an intermediate momentum $m\tns$ is computed by approximating explicitly the pressure in~\eqref{se:adim_m}, then it is updated when the pressure $P\tnp$ is known.
More precisely, the following steps are performed for each phase:
\begin{enumerate}[label=\bf\roman*)]
\item the density equation~\eqref{se:adim_d} is solved by using $u_i\tn$;
\item the volume fraction equation~\eqref{se:adim_a} is solved by using $u_I\tn$ (only for the phase $i$, then $\alpha_{i*}=1 - \alpha_i$);
\item the intermediate momentum $(\alpha m)\tns= ( \alpha \rho_i)\tnp u_i\tns $ is computed solving
\begin{equation}
\label{e:m_tns}
\frac{(\alpha_i m_i)\tns - (\alpha_i m_i)\tn}{\Delta t}  + 
  \dpar{\left[(\alpha_i m_i)\tns u_i\tn + \alpha_i\tnp P_i\tn \right]}{x} =
\;P_I\tn \dpar{\alpha_i\tnp}{x} \; ;
\end{equation}
\item the pressure-equation~\eqref{se:adim_P} is solved by treating implicitly the velocity divergence (which is mandatory to avoid too severe restrictions on the time step) and explicitly the terms related to the speed of sound, that is
\begin{multline}
\label{e:P_tn}
 M_r^2 \alpha_i\tnp \!\! \left[\frac{P_i\tnp - P_i\tn}{\Delta t}  +  u_i\tns \dpar{P_i\tnp}{x} \right] 
 + \left( M_r^2 \rho_i c_i^2 + \kappa_i \right)\tn \alpha_i\tnp \dpar{u_i\tnp}{x}\\
 = \left( M_r^2 \rho_i c_{i,I}^2 + \kappa_i \right)\tn  (u_I - u_i)\tns \dpar{\alpha_i\tnp}{x} \;
\end{multline}
\item the momentum is updated by solving the difference between \eqref{se:adim_m} and \eqref{e:m_tns}:
\begin{multline}
\label{e:m_tnp}
\frac{(\alpha_i m_i)\tnp - (\alpha_i m_i)\tns}{\Delta t} 
 + \dpar{\left[ \left((\alpha_i m_i)\tnp - (\alpha_i m_i)\tns \right) u_i\tn\right]}{x}\\
 + \dpar{\left[\alpha_i\tnp (P_i\tnp - P_i\tn)\right]}{x}
 = (P_I\tnp - P_I\tn) \dpar{\alpha_i\tnp}{x} \, .
\end{multline}
\end{enumerate}

\subsubsection{The staggered spatial discretization.}
We use a pretty standard first-order, finite-volume discretization with Rusanov fluxes, so we highlight here only the peculiar features of the proposed scheme.
Given a 1D domain, we divide it in $N_j^s$ scalar cells of uniform width $\Delta x_j$ and $N_j^v = N_j^s+1$ cells for the vectorial variables, whose width $\Delta x\jph$ is the equal to $\Delta x_j$ for the internal cells, except for the first and the last cells which have width $\Delta x_j/2$.

The equations for $\alpha$, $\alpha \rho$, $P$ are solved over the scalar (node-centered) cells $\mathcal{C}_j$, while the equations for $\alpha m$ are solved over the vector cells $\mathcal{C}\jph$.
When a mapping from the vectorial to the scalar cells, or vice versa, is required, we use a weighted average according to the cell sizes, which, in this preliminary work, reduces to a simple mean as we use uniform grids.
We denote the spatially discrete values by a further subscript $j$ or $j+\frac{1}{2}$, e.g., $(\alpha_i)_j\tn$ is the volume fraction of phase $i$ at cell $j$ at time step $n$.
The staggered configurations facilitates the computation of some integral terms: for instance, the velocity on the face between the cells $\mathcal{C}_j$ and  $\mathcal{C}_{j+1}$ is unequivocally $u\jph$. Similarly, we can easily discretize the pressure gradient over the cell $\mathcal{C}\jph$ through a central difference between $P_{j+1}$ and $P_j$.

Particular care is mandatory while discretizing the non-conservative terms, which involves the gradient of $\alpha$. As known, 
a superficial discretization of these products may lead to the onset of oscillations across interfaces separating fluids with different material properties~\cite{Abgrall2010}.
In compressible multiphase flows, a guideline is provided by the so-called Abgrall's criterion (or pressure non-disturbance condition), which states that ``a two-phase flow, uniform in pressure and velocity must remain uniform on the same variables during its temporal evolution''~\cite{Abgrall1996}.
So, applying the derived discretization to a uniform flow, we can obtain an oscillation-free discretization of the non-conservative terms. This procedure clearly depends on the adopted discretization scheme, but it has been already applied to BN-type models in previous works of Saurel et al.~\cite{Saurel1999,Saurel2017}.

According to the previous remarks, the discrete equations solve at each time step are
\begin{enumerate}[label=\bf\roman*)]
\item  $ (\alpha_i \rho_i)_j\tnp = (\alpha_i \rho_i)_j\tn  - \frac{\Delta t}{\Delta x_j} 
\left[F\jph^{\mathsf{rus},\rho}  
 - F\jmh^{\mathsf{rus},\rho}  \right] \qquad $  with \hfill \inlineeqno\small
\begin{equation*}
F\jph^{\mathsf{rus},\rho} = 
\frac{1}{2} \left[ (\alpha_i \rho_i)_{j+1}\tnp +(\alpha_i \rho_i)_j\tnp \right] (u_i)\jph\tn -
\frac{1}{2} \left[ (\alpha_i \rho_i)_{j+1}\tnp \! - (\alpha_i \rho_i)_j\tnp \right]\!\! \vert(u_i)\jph\tn\vert 
\end{equation*}
\normalsize\vspace*{-1.8mm}
\item \label{step:alpha}
$(\alpha_i)_j\tnp = (\alpha_i)_j\tn  - \frac{\Delta t}{\Delta x_j}  \mathsf{H}_u(\alpha_i\tnp, u_I\tn)_j \quad$\hfill \inlineeqno

\vspace*{2mm}
where $\mathsf{H}_u$ is the discretization of the non-conservative term. It is obtained by combining this equation with the one at step \textbf{i)}: starting from uniform density and velocity, we impose that the density remains constant, obtaining \small
\begin{multline*}
\mathsf{H}_u(\alpha_i\tnp \!\!\!, u_I\tn)_j = \frac{1}{2} 
\left\lbrace 
\left[ (\alpha_i)_{j+1}\tnp - (\alpha_i)_{j-1}\tnp \right] (u_I)_j\tn  \right.\\
\left.
 - \vert(u_I)_j\tn\vert \left[ (\alpha_i )_{j+1}\tnp \! - 2(\alpha_i)_j\tnp + (\alpha_i)_{j-1}\tnp \right]
\right\rbrace 
\end{multline*}\normalsize
\item \label{step:m} $\begin{aligned}[t]
(\alpha_i m_i)\jph\tns &= (\alpha_i m_i)\jph\tn
  -\tfrac{\Delta t} {\Delta x\jph} \Big\lbrace  \left[F_{j+1}^{\mathsf{rus},m}  
 - F_j^{\mathsf{rus},m}  \right] \\
 +&  \left[ (\alpha_i)_{j+1}\tnp (P_i)_{j+1}\tn 
      - (\alpha_i)_{j}\tnp(P_i)_j\tn  \right]  
  - \mathsf{H}_P(\alpha_i\tnp, P_I\tn, u_i\tns)\jph\tns   \Big\rbrace \;\;\; \inlineeqno
\end{aligned} $\vspace*{4mm}

where $F_{j}^{\mathsf{rus},m}$ is the standard Rusanov flux for the conservative variables $(\alpha_i m_i)\tns$ and the convective velocities $(u_i)\tn$, while the approximation $(\mathsf{H}_P)\jph\tns$ of the non-conservative term is obtained by imposing the Abgrall's criterion: \small
\begin{equation*}
(\mathsf{H}_P)\jph\tns =
  (P_I)\tn\jph \left[(\alpha_i)_{j+1}\tnp  - (\alpha_i)_{j}\tnp \right]
 + K^\mathrm{vol}\jph (u_i)\jph\tns
 \left( F_{j+1}^{\mathsf{rus},\rho, \bar{u}} - F_{j}^{\mathsf{rus},\rho, \bar{u}} \right)
\end{equation*} \normalsize
where $K^\mathrm{vol}\jph=[1- \Delta x\jph\big/ \Delta x_j]$ and $F_{j+1}^{\mathsf{rus},\rho, \bar{u}}$ is defined as in (9) but between the \textit{mapped densities}  $(\alpha_i \rho_i)\jph$ and $(\alpha_i \rho_i)\jmh$.
The last term of $\mathsf{H}_P$ results from the mapping of $(\alpha_i \rho_j)$ at the left-hand side,\footnote{
The discretization of $\mathsf{H}_P$ is obtained by assuming uniform $(u_i)\tn$ and $(P_i)\tn$ in (11) and substituting into its left-hand side  $(\alpha_i m_i)\jph\tns=(\alpha_i \rho_i)\jph\tnp$ the discretization of $(\alpha_i \rho_i)\jph\tnp$ from (9).} but it is non-null only at the vector boundary cells, where $\Delta x\jph \neq \Delta x_j$.\vspace*{2mm}
\item \label{step:P}$\begin{aligned}[t]
M_r^2 (\alpha_i)_j\tnp  &\left[ (P_i)_j\tnp -(P_i)_j\tn \right]\tfrac{\Delta x_j }{\Delta t} 
= \\
&-  \tfrac{M_r^2(\alpha_i)_j\tnp }{2} \Big\lbrace \left[ (P_i)_{j+1}\tnp -(P_i)_j\tnp    \right]
     \left[ (u_i)\jph\tns - \vert(u_i)\jph\tns\vert \right]   \\
   &\qquad\qquad\quad\,+ \left[ (P_i)_{j}\tnp   -(P_i)_{j-1}\tnp \right] 
     \left[ (u_i)\jmh\tns + \vert(u_i)\jmh\tns\vert \right] \Big\rbrace \\
&- \left[ M_r^2 (\rho_i)_j\tn ((c_i)_j\tn)^2 \;\,\,+ (\kappa_i)_j\tn \right]
 (\alpha_i)_j\tnp  \left[ (u_i)\jph\tnp - (u_i)\jmh\tnp \right] \\
&+ \left[ M_r^2 (\rho_i)_j\tn  ((c_{i,I})_j\tn)^2 + (\kappa_i)_j\tn \right]
   \mathsf{H}_{u} \left( (u_I\tns- u_i\tns), \alpha_i\tnp \right)_j \qquad\quad \inlineeqno 
\end{aligned}$ \vspace*{4mm}

which has been derived from \eqref{e:P_tn} re-writing $u\tns \dpar{P\tnp}{x} = \dpar{P\tnp u\tns}{x} - P\tns\dpar{u\tns}{x}$ and using the Rusanov approximation.
For analogy with the density equation, the non-conservative term is discretized by the same operator $\mathsf{H}_{u}$ defined in step~\ref{step:alpha}.
\item \label{step:mnp}we observe that if in Eq.~\eqref{e:m_tnp} we define  $\delta \alpha_i m_i = (\alpha_i \rho_i)\tnp (u_i\tnp - u_i\tns)$ and  $\delta P = P\tnp - P\tn$, it has the same shape of \eqref{e:m_tns}. Therefore we obtain the same discretization as in step~\ref{step:m}.
\end{enumerate}

We highlight that the velocity in the second right-hand side term of (12) has to be discretized implicitly, but at step~\ref{step:P} the velocities $(u_i)\tnp$ are still unknown.
However, we can derive an approximation from the expression of step~\ref{step:mnp}, discharging the differences in the convective term. Thus, we use
\begin{multline}
(u_i)\jph\tnp =
(u_i)\jph\tns
+ \tfrac{\Delta t} {\Delta x\jph}
  \Big\lbrace - \left[ (\alpha_i)_{j+1}\tnp (\delta P_i)_{j+1}
       - (\alpha_i)_{j}\tnp(\delta P_i)_j\tnp \right]  \\
  + (\delta P_I)\tnp\jph  \left[ (\alpha_i)_{j+1}\tnp  - (\alpha_i)_{j}\tnp \right] \Big\rbrace
  \Big / {(\alpha_i \rho_i)\jph\tnp}\;.
\end{multline}
However, the previous expression contains the interfacial velocity $P_I\tnp$, which depends on the pressure at time $t\tnp$ of both phases.
Consequently, the use of (13) in (12), i.e. the implicit discretization of the acoustic part, makes the solution of the pressure equations coupled among all phases.

\section{Results and Discussion}
\label{s:results}
We have tested the proposed numerical method on some simplified problems to verify its correctness.
In the presented tests, we use the stiffened gas equation of state, which describes molecular agitation and
repulsive effects in a simplified way, but facilitates the analytical solution of some simple problems~\cite{Haller2003}.

\begin{table}\noindent
\begin{minipage}{0.4\linewidth}\centering
\begin{tabular}{lll}
\hline
                 &  $\gamma\quad\;$ &  $P_\infty~(\mathrm{Pa})$\\
\hline\rule{0pt}{12pt}%
\textsf{Phase 1}$\;\;$  &  4.4     &  $6.0 \, 10^8$ \\
\textsf{Phase 2}        &  1.4     &   0.0 \\
\hline
\multicolumn{3}{l}{\rule{0pt}{12pt}%
\textbf{a)} Fluids of tests 1 and 3.}
\end{tabular}
\end{minipage}\hspace{1cm}
\begin{minipage}{0.4\linewidth}\centering
\begin{tabular}{llll}
\hline
                 & $c_{p}~(\mathrm{J kg^{-1} K^{-1}})\;\;$ & $\gamma\quad\;$ &  $P_\infty~(\mathrm{Pa})$ \\
\hline\rule{0pt}{12pt}%
\textsf{Phase 1}$\;\;$ & 4186      &  2.8     &  $8.5 \, 10^8$ \\
\textsf{Phase 2} & 1004      &  1.4     &   0.0 \\
\hline
\multicolumn{4}{l}{\rule{0pt}{12pt}%
\textbf{b)} Fluids of test 2.}
\end{tabular}
\end{minipage}
\caption{Thermodynamic parameters of the two fluids. $c_{p}$ is the isobaric specific heat capacity, $\gamma$ is the ratio between the isobaric and the isochoric specific heat capacities, and $P_\infty$ is a parameter of the stiffened gas model.
Values are taken from~\cite{Saurel2017,Pandare2018}.}
\label{t:tmd}
\end{table}

\begin{figure}
\includegraphics[width=\textwidth]{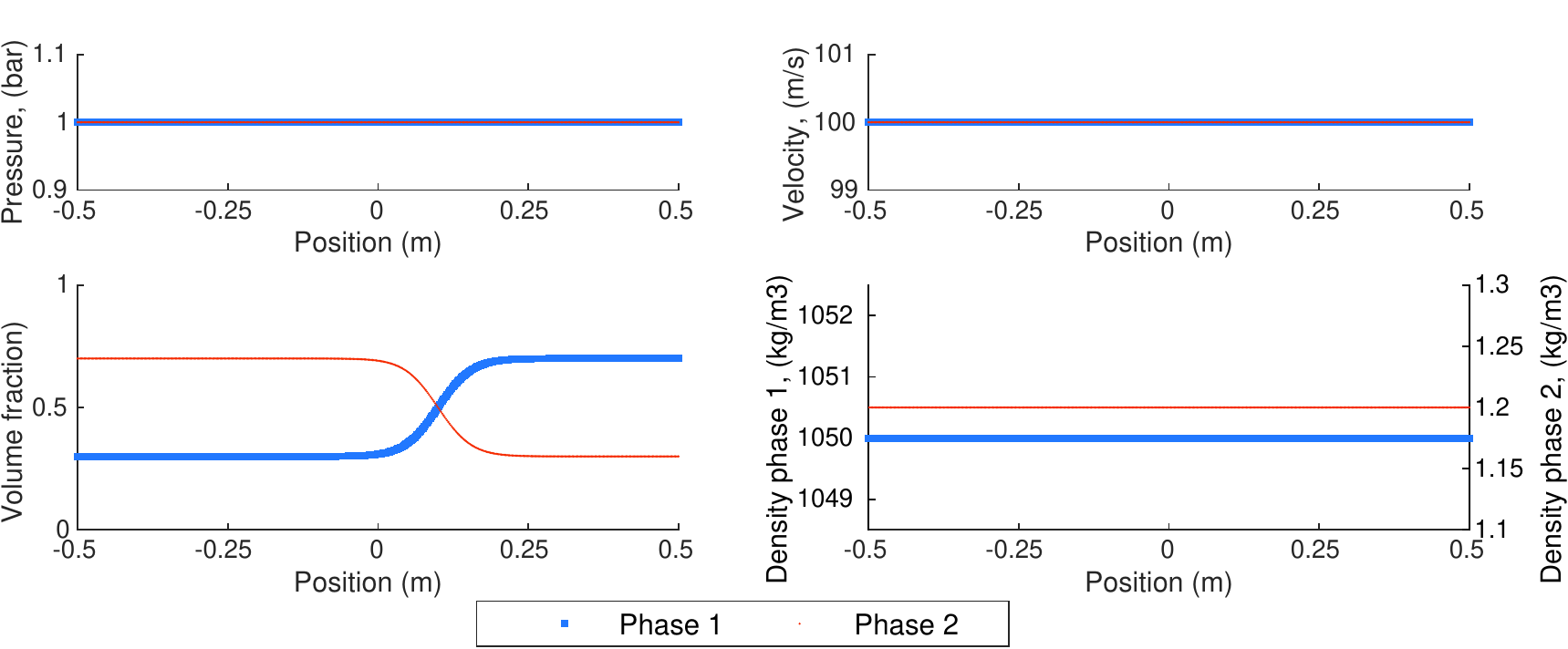}
\caption{Test 1: volume fraction transport in a uniform pressure and velocity field, solution at the final time $t^\mathrm{F}=1~\mathrm{ms}$. The domain is $x=-0.5~\mathrm{m}\leq x \leq 0.5 ~\mathrm{m}$, discretized with $N_j^s=500$ scalar cells.
Initial conditions are: $u^0=100~\mathrm{m/s}$, $P^0=10^5~\mathrm{Pa}$, $\rho_1^0 = 1050~\mathrm{kg/m^3}$ and $\rho_2^0=1.2~\mathrm{kg/m^3}$.
The volume fraction profile is initially centered at $x=0$, with $\alpha_1=0.3$ at left and $\alpha_1=0.7$ at right.}
\label{f:uni}
\end{figure}

\paragraph{Test 1.}
To verify the numerical verification of Abgrall's criterion and the stability at CFL$>1$, the first test is a simple transport of a volume fraction variation in a uniform pressure and velocity field.
The fluids and the initial conditions are described in Table~\ref{t:tmd} (left) and Fig.~\ref{f:uni}.
The final time is $1~\mathrm{ms}$, divided in 80 time steps.
The resulting acoustic CFL conditions (i.e., considering $u+c$) are 10 for fluid 1 and 2.7 for fluid 2.
The initial conditions are correctly preserved and no spurious oscillations are generated, as demonstrated in Fig.~\ref{f:uni}.

\begin{figure}
\includegraphics[width=\textwidth]{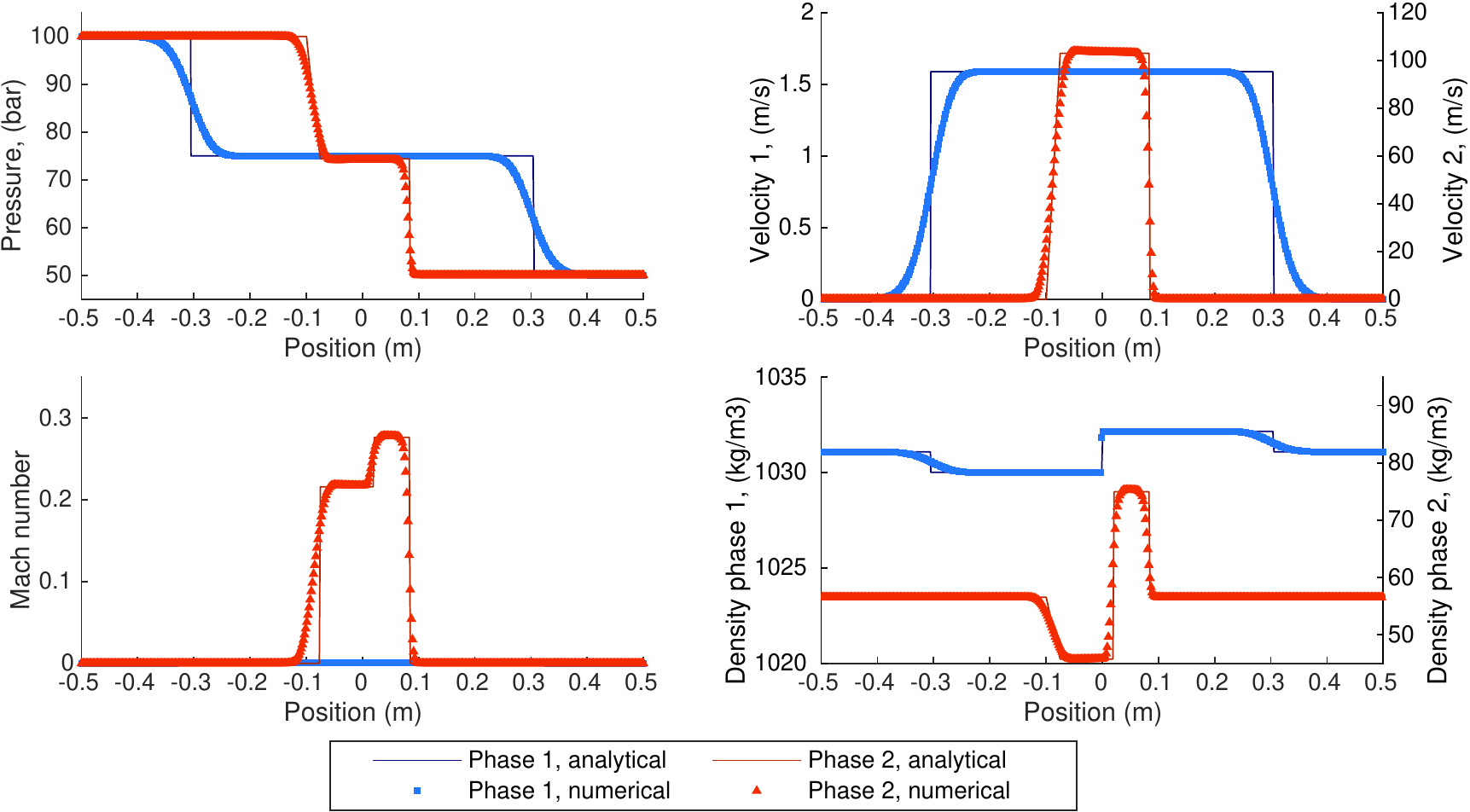}
\caption{Test 2: water/air shock-tube with no phase mixing, solution at the final time $t^\mathrm{F}=0.2~\mathrm{ms}$, after 100 time steps. The grid spacing is $\Delta x_j = 0.002$ as in the other tests.
The initial pressures are $P_\mathrm{L}^0=100~\mathrm{bar}$ (left) and $P_\mathrm{R}^0=50~\mathrm{bar}$ (right), while the temperature is $T^0=308.15~\mathrm{K}$, uniform. The volume fraction is $\alpha=0.5$ everywhere. The numerical solution is compared to the analytical one for each phase.}
\label{f:noMix}
\end{figure}

\paragraph{Test 2.} The second test is a water/air shock-tube with uniform volume fraction $\alpha_1=0.5$. Initially, the left and right chambers have different pressures. The difference in pressure is moderate, in order to avoid excessively high Mach numbers. Because of the absence of relaxation terms, each fluid evolves as a single phase, so it is possible to compute the analytical solution, given the fluid parameters in Table~\ref{t:tmd} and initial conditions in Fig.~\ref{f:noMix}.
A good agreement is achieved, even if the rarefaction fans are smeared, but we can expect it since the discretization is only first-order accurate. We highlight that the pressure variation in the liquid (phase 1) is an acoustic wave and it moves at the speed of sound $c_1\approx1525~\mathrm{m/s}$. Such a high propagation speed makes the capability to implicitly treat acoustics of paramount importance to avoid instabilities or an excessively small time step.

\paragraph{Test 3.} Finally, we reproduce a multiphase test for which the results without relaxation terms are available in~\cite{Saurel2017}. The fluids are the same as in Test 1, but now a pressure difference between the left and right state is also imposed. We can observe a good agreement with results in~\cite{Saurel2017}, expect for the pressure profile of phase 1.
This difference can be explained by the fact that they use a simplified Riemann solver according to which no pressure wave is present in this phase.
Furthermore, we notice a strange behavior of the numerical solution across the interface: in each phase, one point is not aligned with the neighboring ones. We believe that it could be generated by the mapping from the scalar to the vector grid and that it could be smooth down by adding the relaxation terms, however we will further investigate this issue.

\begin{figure}
\includegraphics[width=\textwidth]{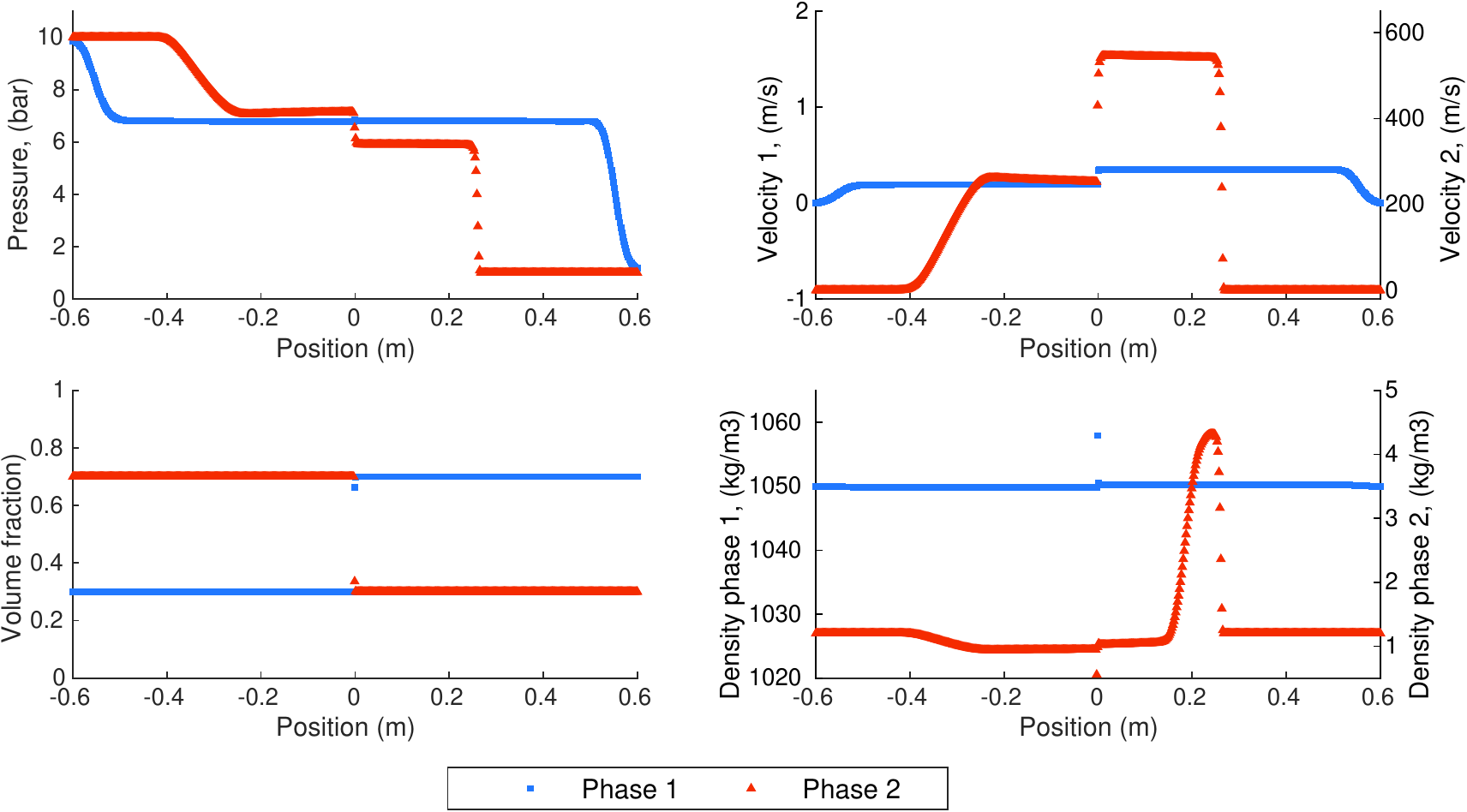}
\caption{Test 3: ``Smooth shock tube test case'' as in~\cite{Saurel2017},
solution at the final time $t^\mathrm{F}=0.35~\mathrm{ms}$, after 700 time steps. The initial pressures are $P_\mathrm{L}^0=10~\mathrm{bar}$ (left) and $P_\mathrm{R}^0=1~\mathrm{bar}$ (right) and the velocity is $0$ for both phases.
The densities $\rho_1^0 = 1050~\mathrm{kg/m^3}$ and $\rho_2^0=1.2~\mathrm{kg/m^3}$ are uniform along the domain.
}
\end{figure}

\section{Conclusions}
\label{s:concl}
In this work, we have presented a semi-implicit method for multiphase flows based on the Baer and Nunziato model, which avoids severe time step restrictions at low-Mach regimes.
We have proposed a pressure-based method, which, thanks to a special scaling of the pressure, recovers in the limit $M\Ref \rightarrow 0$ the incompressible formulation of the governing equations.
Moreover, we have taken particular care in the discretization of the non-conservative terms, in order to avoid spurious oscillations across the multi-material interfaces.

The preliminary tests have confirmed the implicit treatment of acoustic and the fulfillment of the Abgrall's criterion.
These results indicate the applicability of the propose approach and allow us to pass to the next step of the long-term work.
Currently, we are working on the inclusion of the relaxation terms.
Future steps are the implementation of more accurate thermodynamic models, of second- and high-order discretization.
The flexibility of the BN-type model will allow also the investigation of flows containing more than 2 phases.

\subsubsection{Acknowledgments.} This publication has been produced with support from the NCCS Centre, performed under the Norwegian research program Centres for Environment\hyp{}friendly Energy Research (FME). The authors acknowledge the following partners for their contributions: Aker Solutions, Ansaldo Energia, CoorsTek Membrane Sciences, Emgs, Equinor, Gassco, Krohne, Larvik Shipping, Norcem, Norwegian Oil and Gas, Quad Geometrics, Shell, Total, V\aa r Energi, and the Research Council of Norway (257579/E20).

%
%

\bibliographystyle{spphys}
\bibliography{Biblio_ReAbgrall}

\end{document}